\documentclass[fleqn,12pt,twoside]{article}
\usepackage{epsfig}
\usepackage{espcrc1}
\usepackage[figuresright]{rotating}

\newcommand{\AmS}{{\protect\the\textfont2
  A\kern-.1667em\lower.5ex\hbox{M}\kern-.125emS}}
\makeatletter
\newbox\slashbox \setbox\slashbox=\hbox{\large$/$}
\def\pslash#1{\setbox\@tempboxa=\hbox{$#1$}
  \@tempdima=0.5\wd\slashbox \advance\@tempdima 0.5\wd\@tempboxa
  \copy\slashbox \kern-\@tempdima \box\@tempboxa}

\title{Lowest eigenvalues of the Dirac operator for two color
QCD at finite density\thanks{Supported in part by FWF project P14435-TPH.}}

\author{Elmar Bittner\address{Atominstitut, Technische Universit\"at Wien,
        A-1040 Vienna, Austria}, Maria-Paola Lombardo\address{INFN,
        Sezione di Padova, e Gruppo Collegato di Trento, Italy},
        Harald Markum$^{\rm a}$, and Rainer Pullirsch$^{\rm a}$}

\begin{document}

\maketitle

\begin{abstract}
  We investigate the eigenvalue spectrum of the staggered Dirac matrix
  in full QCD with two colors and finite chemical potential.
  Along the strong-coupling axis up to the temperature phase transition,
  the low-lying Dirac spectrum is well described by random matrix theory
  (RMT) and exhibits universal behavior. The situation is 
  discussed in the chirally symmetric phase and no universality is seen
  for the microscopic spectral density.
\vspace*{5mm}
\end{abstract}

We have continued our investigations~\cite{Bitt00a,Bitt00}
of the small eigenvalues in the whole phase diagram. The
Banks-Casher formula relates the Dirac eigenvalue density
$\rho(\lambda)$ at $\lambda=0$ to the chiral condensate,
$ \Sigma \equiv |\langle \bar{\psi} \psi \rangle| =
 \lim_{\varepsilon\to 0}\lim_{V\to\infty} \pi\rho (\varepsilon)/V$.
The microscopic spectral density,
$ \rho_s (z) = \lim_{V\to\infty}
 \rho \left( {z/V\Sigma } \right)/V\Sigma , $
should be given by the appropriate prediction of RMT.
In the presence of a chemical potential $\mu > 0$, leading to
complex eigenvalues, the situation is more complicated
and $\rho(0)$ can be used as a lower bound for
$\Sigma$.

\begin{figure*}[t]
    \centerline{\hspace{3mm}
              \psfig{figure=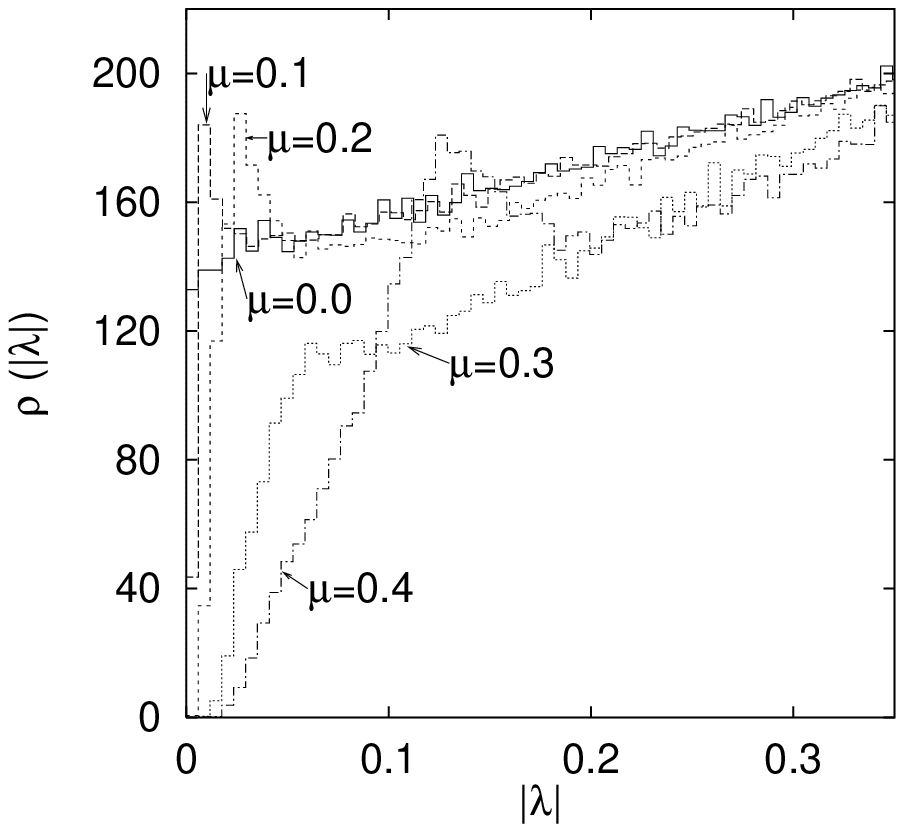,width=5cm}\hspace*{1mm}
              \psfig{figure=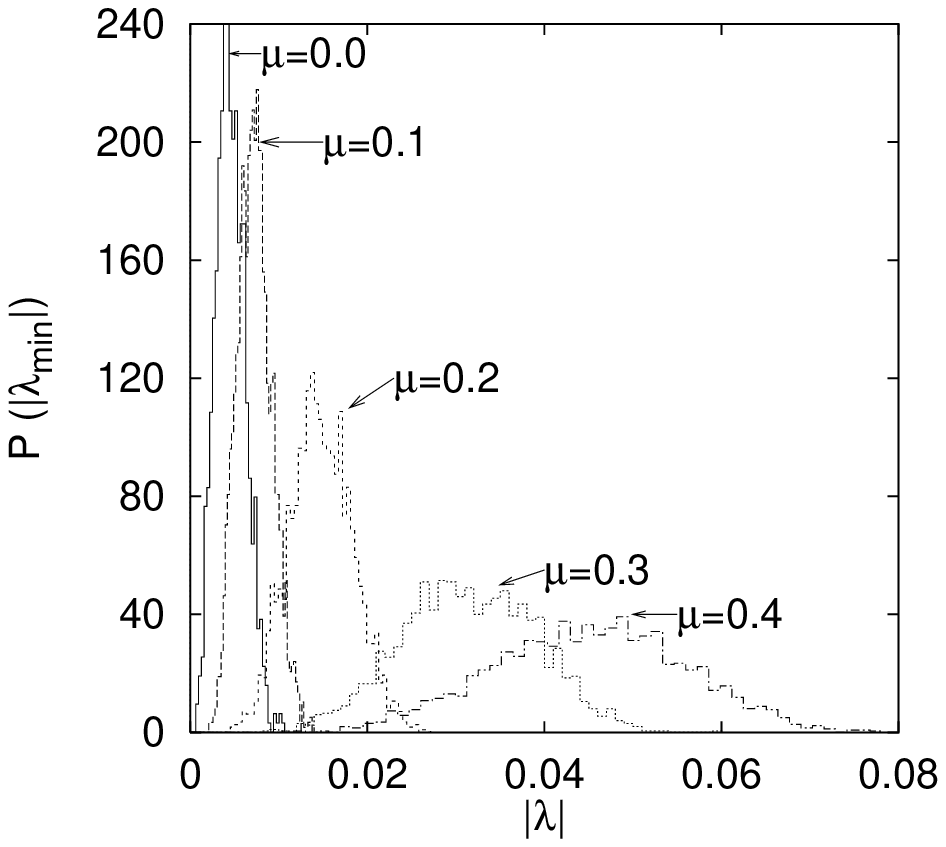,width=5cm}\hspace*{1mm}
              \psfig{figure=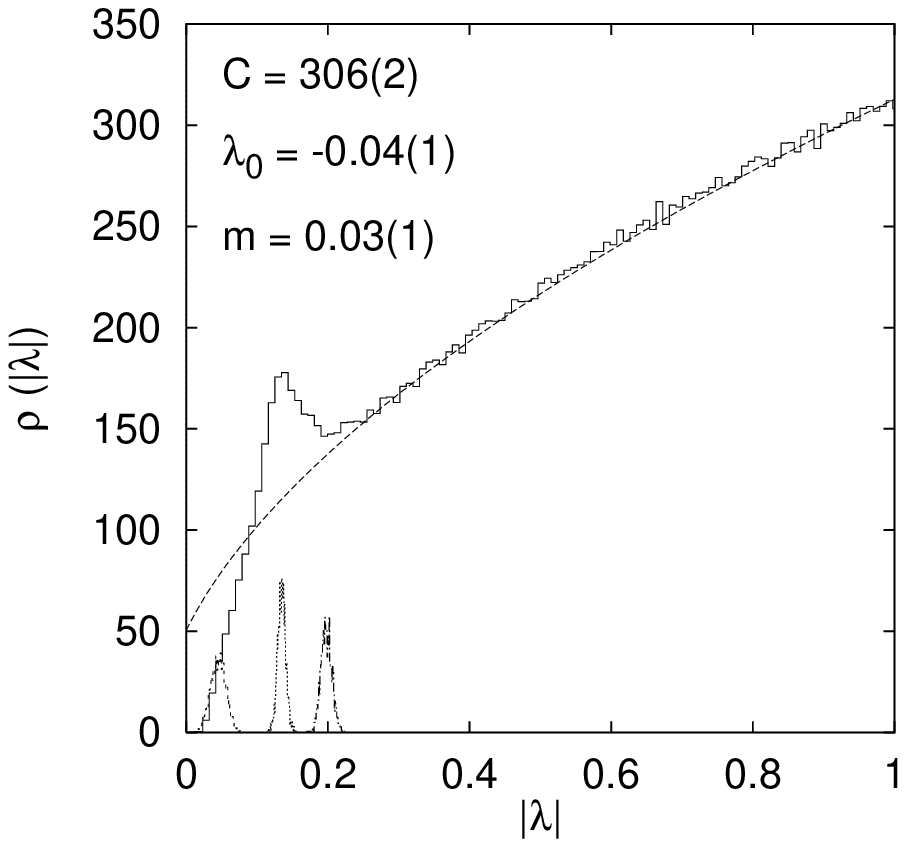,width=5cm}
               }
  \centerline{\hspace{3mm}
              \psfig{figure=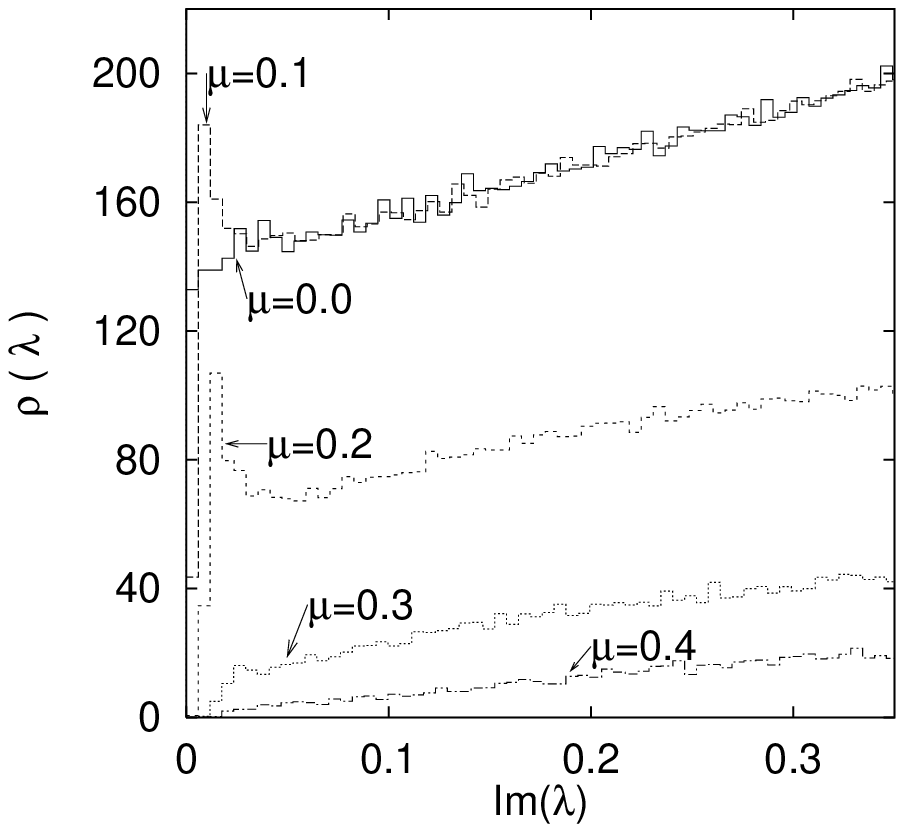,width=5cm}\hspace*{1mm}
              \psfig{figure=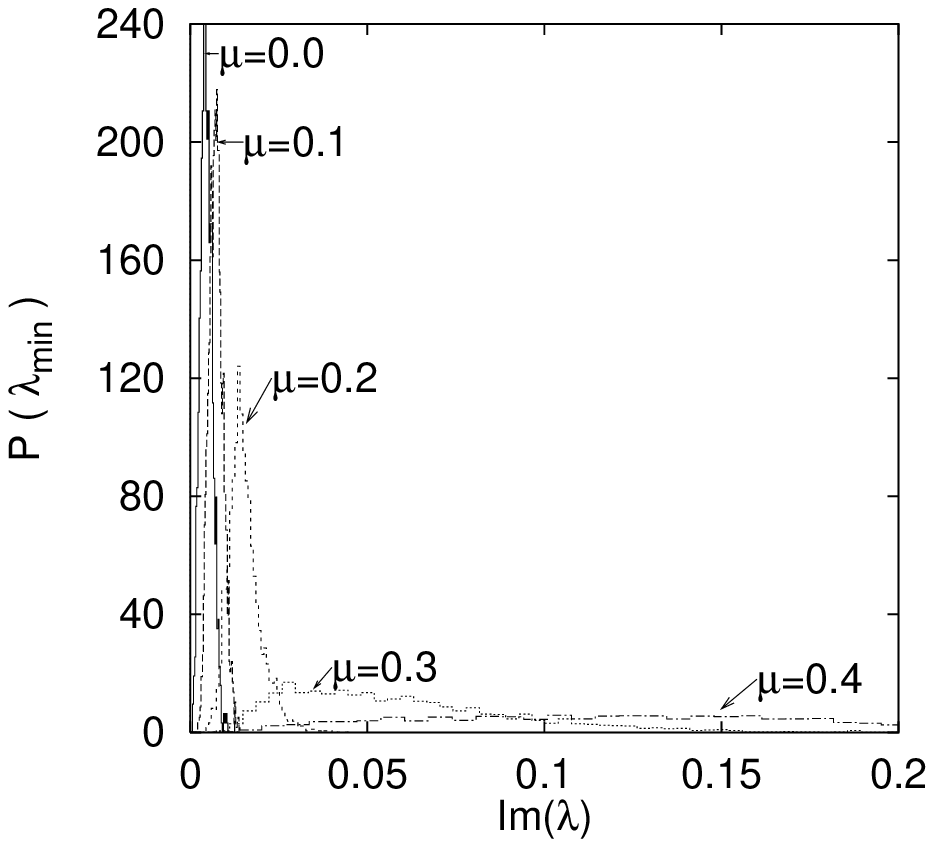,width=5cm}\hspace*{1mm}
              \psfig{figure=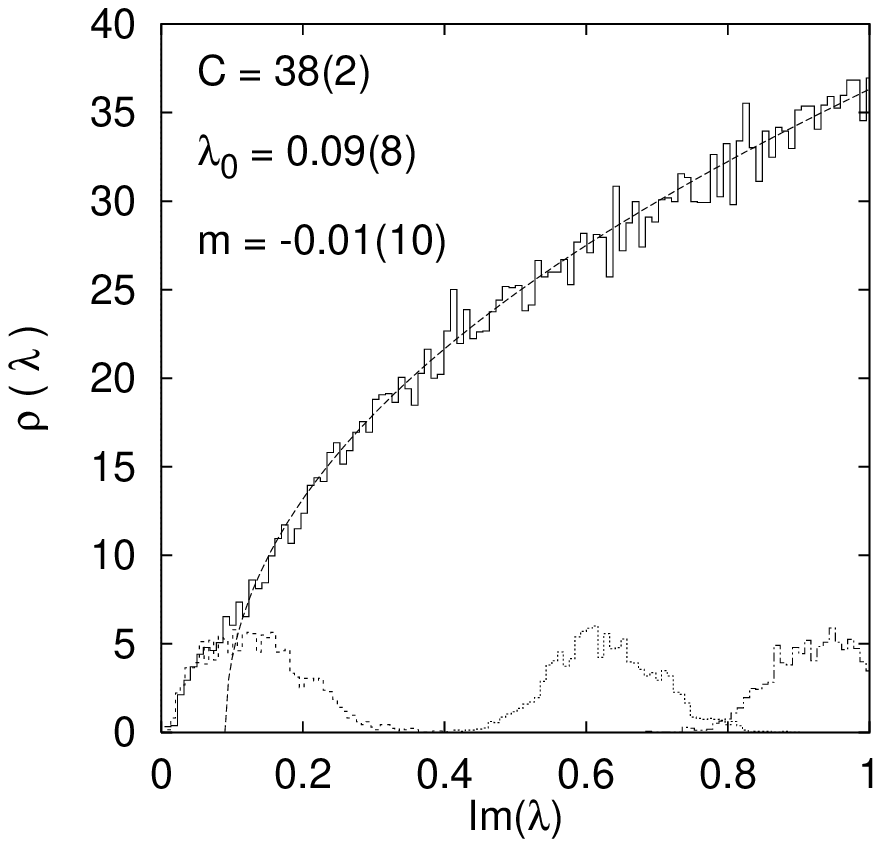,width=5cm}
             }
\vspace*{-10mm}
  \caption{Density $\rho (\lambda)$ of small eigenvalues (left) and
           distribution $P(\lambda_{\min})$ of the smallest eigenvalue
           (center) across the transition of critical chemical
           potential. Fit of the spectral density to $\rho (\lambda) =
           C (\lambda -\lambda_0)^{2m+1/2}$  at $\mu=0.4$ (right) with
           the contribution of the smallest, $11^{{\mbox{th}}}$
           and $21^{{\mbox{st}}}$ eigenvalue inserted. The
           upper plots refer to $\rho(\lambda)$ constructed from the absolute
           value of $\lambda$ while the lower plots result from averages
           over $\cal{R}\rm{e}(\lambda)$.
           }
\vspace{-5mm}
  \label{fig4}
\end{figure*}

We present results for the density of the small eigenvalues in the left
plots of
Fig.~\ref{fig4} for two-color QCD with staggered fermions on a $6^4$
lattice around the critical chemical potential $\mu_c \approx 0.3$
keeping $\beta =1.3$ fixed~\cite{Hand99}. Since the eigenvalues
move into the complex plane for $\mu > 0$, $\rho(|\lambda|)$ was constructed
from the absolute value for $|\lambda|$ small. Alternatively,
a band of width $\epsilon
= 0.015$ parallel to the imaginary axis was considered to construct
$\rho(y)$, i.e. $\rho(y)\equiv\int_{-\epsilon}^\epsilon
dx\,\rho(x,y)$, where $\rho(x,y)$ is the density of the complex
eigenvalues $\lambda=x+iy$. Both results are compared in the upper and
lower plot.

The distribution of the lowest eigenvalue is displayed in the center plots
of Fig.~\ref{fig4}.  In the confinement it was shown for $\mu=0$ that
both the microscopic spectral density $\rho_s(z)$ and the distribution
$P(\lambda_{\rm min})$ of the smallest eigenvalue agree with the RMT
predictions of the chiral symplectic ensemble~\cite{Berb98}. The quality
of the data from at least 2100 configurations at $\mu\neq 0$ was not sufficient
for a reliable fit to a trial function for $P(\lambda_{\min})$.

Nevertheless, the quasi-zero modes which are responsible for the chiral
condensate $\Sigma \neq 0$ build up when we cross from the plasma
into the confined phase. Both $\rho(\lambda)$
and $P(\lambda_{\min})$ plotted with varying $\mu$ on identical scales
can serve as an indicator for the phase transition.

In the right plots of Fig.~\ref{fig4} we discuss the
spectrum in the quark-gluon plasma. From RMT a functional form of
$\rho (\lambda) = C (\lambda -\lambda_0)^{2m+1/2}$ is expected at
the onset of the eigenvalue density~\cite{Bowi91}. 
We tried a fit and obtained
consistency with universality class $m=0$~\cite{Lang99,Bitt00}.

In summary, we studied universality concerning the low-lying spectra of the
Dirac operator for two-color QCD with chemical potential. We could not obtain
a relation for $\rho_s(z)$ in $0<\mu<\mu_c$, not only because of our data but 
also in lack of an analytic result for non-Hermitian RMT.
In the phase where chiral symmetry is restored one has to rely on ordinary RMT.
Here we find universal behavior only of the macroscopic density $\rho(\lambda)$.

\end{document}